%% file: modulock.tex
\documentclass[10pt,letterpaper,twocolumn]{article}
\usepackage[10pt,nocopyright]{sigmin}

\usepackage{listings}
\usepackage{color}
\usepackage{epsfig,makecell}    
\usepackage{ifpdf}
\ifpdf
  \usepackage{epstopdf}
\fi                                  %
\usepackage[square,comma,numbers,sort&compress]{natbib}          %
\usepackage[hyphens]{url}                                         %
\usepackage[colorlinks]{hyperref}                      %
\usepackage[usenames,dvipsnames]{xcolor}                          %
\hypersetup{citecolor=blue,linkcolor=blue}                        %
\usepackage{amsmath,amsopn,amssymb,amsthm}                        %
\usepackage{thmtools}
\usepackage[toc,page]{appendix}
\usepackage{subfigure}
\usepackage{endnotes,microtype,xspace,fancyvrb,multirow} %
\usepackage{booktabs}                                             %
\usepackage{array,underscore,relsize}                             %
\usepackage[T1]{fontenc}                                          %
\usepackage{times}                                                %
\usepackage{fancyhdr,lastpage}                                    %
\usepackage{enumitem}                                             %
\usepackage[export]{adjustbox}                                    %
\usepackage{breakurl}                                             %
\usepackage{pifont}                                               %
\usepackage{tablefootnote}                                        %
\usepackage[linesnumbered,vlined,ruled,commentsnumbered]{algorithm2e}
\DontPrintSemicolon

\SetCommentSty{mycommfont}

\usepackage{amssymb}
\usepackage[normalem]{ulem}
\renewcommand{\headrulewidth}{0pt}                                %
\usepackage[hang,flushmargin]{footmisc}
\usepackage{textcomp}
\usepackage{graphicx}

\usepackage[compact]{titlesec}
\titleformat*{\section}{\large\bfseries}
\titleformat*{\subsection}{\normalsize\bfseries}
\titleformat*{\subsubsection}{\normalsize\bfseries}
\titlespacing*{\section}{0pt}{*2}{*1}
\titlespacing*{\subsection}{0pt}{*1.5}{*0.8}

\definecolor{mygreen}{rgb}{0,0.6,0}  
\definecolor{mygray}{rgb}{0.5,0.5,0.5}  
\definecolor{mymauve}{rgb}{0.58,0,0.82}

\hypersetup{
  colorlinks,
  linkcolor={red!50!black},
  citecolor={blue!50!black},
  urlcolor={blue!80!black}
}

\input{cmds.tex}

\usepackage{balance}
\usepackage{upgreek}

\begin{document}

\pagestyle{fancy}                                                 %
\fancyhf{}                                                        %
\renewcommand{\headrulewidth}{0pt}                                %
\cfoot{\thepage}                                                  %

\title{\Large \bf {Towards Lock Modularization for Heterogeneous Environments}}

\author{
    Hanze Zhang\;\;\;\;
    Rong Chen\;\;\;\;
    Haibo Chen\\[5pt]
 \normalsize{{Institute of Parallel and Distributed Systems, Shanghai Jiao Tong University}} \\ [15pt]
}

\maketitle
\frenchspacing

\input{abs.tex}
\input{intro.tex}
\input{problem.tex}
\input{proposal.tex}
\input{discussion.tex}

\balance

\small{
  \bibliographystyle{plain}
  \bibliography{bib/dblp,bib/misc}
}

\clearpage

\end{document}

%% file: cmds.tex
\newcommand{\squishlist}{
  \begin{list}{$\bullet$}{
    \setlength{\itemsep}{0pt}
    \setlength{\parsep}{3pt}
    \setlength{\topsep}{3pt}
    \setlength{\partopsep}{0pt}
    \setlength{\leftmargin}{1.5em}
    \setlength{\labelwidth}{1em}
    \setlength{\labelsep}{0.5em}
  }
}

\newcommand{\squishend}{
  \end{list}
}

\newcounter{enumctr}
\newcommand{\squishenum}{
  \begin{list}{\arabic{enumctr}.}{
    \usecounter{enumctr}
    \setlength{\itemsep}{0pt}
    \setlength{\parsep}{3pt}
    \setlength{\topsep}{3pt}
    \setlength{\partopsep}{0pt}
    \setlength{\leftmargin}{1.5em}
    \setlength{\labelwidth}{1em}
    \setlength{\labelsep}{0.5em}
  }
}

\usepackage{tikz}

\newcommand{\stitle}[1]{\vspace{1.ex}\noindent{\bf #1}}
\newcommand{\nospacestitle}[1]{\noindent{\bf #1}}

\newcommand{\x}{{$\times$}\xspace}

\let\oldding\ding%
\renewcommand{\ding}[2][1]{\scalebox{#1}{\oldding{#2}}}%

%% file: abs.tex
\begin{abstract}
Modern hardware environments are becoming increasingly heterogeneous,
leading to the emergence of applications specifically designed to 
exploit this heterogeneity. 
Efficiently adopting locks in these applications poses distinct challenges. 
The uneven distribution of resources in such environments 
can create bottlenecks for lock operations, severely hindering application performance. 
Existing solutions are often tailored to 
specific types of hardware, which underutilizes resources 
on other components within heterogeneous environments.

This paper introduces a new design principle: decomposing locks across hardware 
components to fully %
utilize unevenly distributed resources in heterogeneous environments. 
Following this principle, we propose \emph{lock modularization}---a systematic approach 
that decomposes a lock into independent modules and assigns them 
to appropriate hardware components. This approach aligns the resource requirements of 
lock modules with the attributes of specific hardware components, 
maximizing strengths while minimizing weaknesses.

\end{abstract}

%% file: intro.tex
\section{Introduction}

Heterogeneity is an emerging trend in the evolution of both production
and edge hardware architectures~\cite{omnix,molecule,fractos}. 
In the post-Moore era, the performance scaling
of CPUs no longer meets the rapidly increasing demands for
highly parallel and energy-efficient processing. Consequently,
processing units and hardware architectures with domain-specific
attributes are being widely adopted, creating various heterogeneous
environments~\cite{bf3,tofino,dmos,tpuv4,faery}.

As a fundamental synchronization primitive, locks are commonly used in 
applications designed for heterogeneous environments. For instance,
transactional databases empowered by programmable networks and
memory disaggregation~\cite{polardbsrvless,polardbmp,dpustudy}
adopt locks to serialize index updates~\cite{sherman,smart,chime} and
enforce inter-transaction concurrency control~\cite{ford,motor}.
XPU-accelerated applications require locks to coordinate between
algorithm modules~\cite{autoware,cyberrt} and serialize updates to
large-volume structural 
data~\cite{Kai2019GPUlock,Lan2022GPUlock,Gao2020GPUlock}. 
Since lock operations are on the critical path of shared data access,
they often play a crucial role in application performance.

Despite their importance, efficiently adopting locks in heterogeneous
environments is challenging due to the uneven distribution of
processing, memory, and communication resources across hardware components.
Heterogeneous hardware is typically strong in one type of resource
but weak in others to meet specific workload requirements. 
For example, programmable switches offer high packet processing speed 
but have limited memory capacity~\cite{tofino,cisco,broadcom}. 
Consequently, when adapting locks to specific hardware, bottlenecks
in scarce resources often prevent the full utilization of 
advantageous resources, leading to suboptimal performance~\cite{fisslock}.

Existing hardware-specific lock designs improve performance in two main ways: 
1) by customizing lock algorithms to reduce network contention~\cite{smart24} 
or optimize processing resource utilization~\cite{Gao2020GPUlock,libasl}; 
and 2) by caching frequently accessed locks on specific hardware to leverage 
high processing power~\cite{netlock} or low communication costs~\cite{cohort,sherman}.
However, these solutions are usually tailored for individual hardware components 
based on specific observations, preventing them from fully exploiting resources 
across different components in a heterogeneous environment. 

Recent research~\cite{fisslock} explores a new strategy called
lock fission, which effectively utilizes both
programmable switches and servers for efficient lock management.
It decomposes lock operations into two phases, each with distinct
resource requirements that match the attributes 
of switches and servers, respectively. By assigning these phases
to the best-matched hardware components, this approach can accelerate
operations on millions of locks, two orders of magnitude more
than deploying undecomposed locks to the switch~\cite{netlock}.
Unfortunately, this approach is tightly coupled with the attributes
of programmable switches and cannot be directly applied to
other heterogeneous environments.

\stitle{Design principle.}
Inspired by lock fission~\cite{fisslock}, we propose a principle for designing locks
for heterogeneous environments: \emph{decomposing
lock operations and metadata across different hardware components.}
This principle fully exploits unevenly distributed
hardware resources by assigning lock modules to hardware
components with matching resource attributes.

\stitle{Our proposal.}
To facilitate the application of this design principle in diverse heterogeneous
environments, we propose \emph{lock modularization}---a systematic approach
of decomposing locks and assigning lock modules to hardware components.
This approach breaks the lock manager into smallest functional units,
including managers for mode, holders, waiters, and lock granting. 
Developers can easily assign these units to appropriate hardware components
by matching resource requirements with hardware characteristics,
and subsequently merge units assigned to the same component.
We offer preliminary solutions to address correctness and performance 
challenges in adopting this approach, including a grant counting mechanism 
that preserves the linearizability of lock operations and three module 
assignment heuristics to fully exploit the performance benefits of 
modularization. Furthermore, we show the effectiveness of lock 
modularization in two application scenarios---SmartNICs and disaggregated 
memory---where the modularized design 
achieves significantly higher throughput and lower latency than 
monolithic designs.

%% file: problem.tex
\section{Trends, Challenges, and Practices}\label{sec:bg}

\subsection{The Heterogeneous Trend}

\begin{figure}[t]
  \vspace{1mm}
  \begin{minipage}{1.\linewidth}
    \centering\includegraphics[scale=.54]{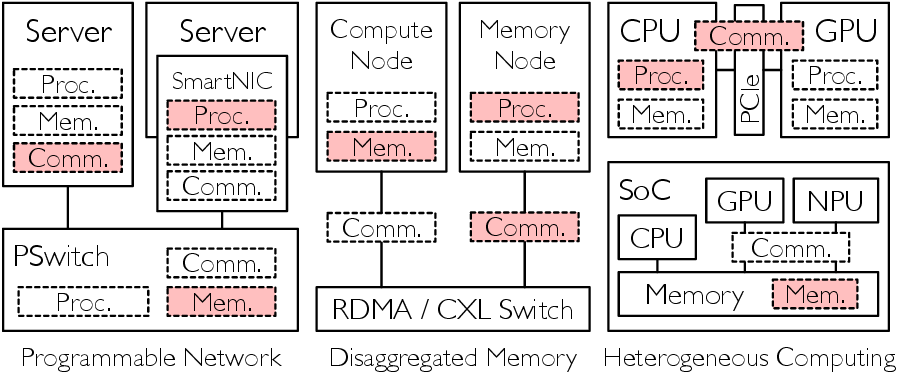}
  \end{minipage} \\[-2pt]
  \begin{minipage}{1.\linewidth}
    \caption{\small{\emph{Heterogeneous environments
    and their attributes in processing (Proc.),
    memory (Mem.), and communication (Comm.) resources.
    Red boxes denote resource bottlenecks in hardware components.
    }}}
    \label{fig:dm}
  \end{minipage}  \\[-5pt]
\end{figure}

To improve performance and energy efficiency when executing diverse tasks, 
both data centers and edge devices are increasingly incorporating 
\emph{heterogeneous} processing units and hardware architectures. 
As shown in Fig.~\ref{fig:dm},
unlike traditional \emph{homogeneous} environments, heterogeneous environments
feature diverse capabilities and limitations that demand 
new system designs to leverage these attributes.
We next discuss these attributes in three typical heterogeneous scenarios.

\stitle{Programmable networks.}
With advancements in high-speed networking, the performance bottleneck
for network-intensive applications is shifting from network transmission 
to packet processing~\cite{netchain,floem,fisslock}. 
In response, programmable network devices, such as 
programmable switches~\cite{tofino,cisco,broadcom} 
and SmartNICs~\cite{stingray,bf2,bf3}, have emerged.
Programmable switches allow customization of packet processing
and forwarding rules. They provide fast packet processing
and short network paths, but their limited memory capacity---typically 
only a few MBs~\cite{tofino}---restricts their ability 
to accelerate large-scale data processing~\cite{fisslock,netcache}.
SmartNICs enhance traditional NICs by incorporating programmable units
to offload simple application tasks. Although these units offer lower 
communication costs than CPUs due to their proximity to the 
network~\cite{dpu3study}, they suffer from either wimpy processing 
power~\cite{dpustudy,dpu3study} or small cache sizes~\cite{dpu3study}, 
which constrains their performance.

\stitle{Disaggregated memory.}
A recent trend in modern data centers is the separation of CPU and memory
resources into disaggregated, network-attached compute nodes (CNs)
and memory nodes (MNs)~\cite{legoos,dmos,firebox}. 
CNs possess strong processing power but limited memory capacity (usually a few GBs), 
whereas MNs have substantial memory capacity but negligible processing power.
This architecture enhances resource utilization and flexibility by 
allowing applications to scale CPU and memory resources independently. 
Applications on disaggregated 
memory~\cite{polardbmp,sherman,smart,chime,motor,ditto} typically 
run compute tasks on CNs, which retrieve and update data stored on MNs
by directly accessing MN memory using technologies such as RDMA and CXL.
Since multiple CNs often access shared data on MNs simultaneously, 
MNs commonly experience a much larger communication burden than CNs~\cite{smart,ditto,chime}.

\stitle{Heterogeneous computing.}
The increasing demand for large-scale and cost-effective computing has led to 
the widespread adoption of heterogeneous processing units (XPUs), including
accelerators such as GPUs, neural processing units 
(NPUs)~\cite{KovacsICVS21,tpuv4}, and domain-specific FPGAs~\cite{faery}.
XPUs can be attached to CPUs as external devices via PCIe, which is common 
in datacenters, or integrated with CPUs as system-on-chips (SoCs) 
in edge devices like smartphones and AI PCs. 
In the former scenario, XPUs may have independent, high-bandwidth memory
with large capacities (up to hundreds of GBs~\cite{b200}). 
However, the PCIe connection and associated software stacks introduce significant overhead 
for communication with CPUs (several microseconds). 
In the latter scenario, the main memory is shared between CPUs and XPUs, 
allowing for lower communication costs (hundreds of nanoseconds) through shared memory. 
However, the memory capacity and bandwidth are constrained.

\subsection{The Lock Adoption Challenge}
\label{sec:prob:challenge}

When designing high-performance applications for heterogeneous 
environments, a significant challenge is the adoption of locks. 
Locking requires adequate processing, memory, 
and communication resources to be efficient. 
However, in heterogeneous environments, these resources are unevenly
distributed across various hardware components. 
Consequently, designing locks for specific hardware would encounter
performance bottlenecks due to limitations in one type of resource,
preventing full exploitation of the advantages in others.

\stitle{Programmable networks.}
In the conventional distributed locking paradigm, clients acquire and
release locks by sending network requests to a lock service running
on a remote server's CPUs. The limited packet processing speed
of CPUs can cause requests to queue up under heavy load, incurring 
significant delays in lock operations~\cite{netlock,fisslock}.
While programmable network devices can help alleviate the lock request 
processing burden on CPUs, their heterogeneous resource attributes
hinder efficient lock design.
SmartNICs, despite having shorter network paths to client machines, 
process lock requests more slowly than CPUs, leading to higher lock
acquisition latencies.
Programmable switches offer higher packet processing speeds, 
but their limited memory capacity 
restricts them to managing only a few thousand locks, which
is inadequate for real-world workloads~\cite{fisslock}.

\stitle{Disaggregated memory.}
In most disaggregated memory applications, locks are stored on MNs
alongside the data they protect~\cite{sherman,smart,chime,ford,motor}.
Clients on CNs acquire and release locks through remote atomic
operations that directly fetch and update lock metadata in MN memory.
Due to the lack of processing power, MNs cannot notify clients 
when a lock is released. 
Therefore, clients must repeatedly check the lock 
states on MNs to determine if they can acquire the lock.
These repeated operations from multiple clients require significant
communication resources on the MN side, which are already limited 
due to remote data access demands.
This saturation of MN communication
resources leads to high network delays, causing significant 
waste of processing and communication resources on CNs.

\stitle{Heterogeneous computing.}
Locking is crucial in heterogeneous computing systems to synchronize
between concurrent modules in autonomous workflows~\cite{autoware,cyberrt} 
or serialize concurrent updates to data structures manipulated by 
XPUs~\cite{Lan2022GPUlock,Gao2020GPUlock}. However, the architectural
attributes of XPUs often hinder efficient lock implementation.
GPUs, with their high parallelism, can generate a massive number
of lock requests that overwhelm the memory controller, impacting 
the memory access performance of other processing units on the same SoC
and wasting their processing resources. 
NPUs typically have fixed functionalities, making it challenging
to implement fully functional locks.

\subsection{Existing Approaches}

Several attempts have been made to optimize locks for specific
heterogeneous hardware. However, these approaches often fail to 
fully exploit heterogeneous resources due to two main reasons.
First, they focus on accommodating the resource attributes of 
a specific hardware component, overlooking the uneven resource 
distribution across different components in heterogeneous environments.
Second, they are tightly coupled with observations from specific hardware,
making it difficult to generalize to other types of hardware.

\stitle{Customizing lock algorithm.}
A common approach involves adjusting the resource requirements of 
lock algorithms to suit the attributes of specific hardware.
For instance, adding backoffs between repeated lock state 
checks can reduce the communication resource requirement of locks
in disaggregated memory environments~\cite{smart24}. 
Combining warp yielding with backoff
can maximize the utilization of GPU processing resources while 
reducing memory access contention~\cite{Gao2020GPUlock}.
Reordering big and small cores in the wait queue can improve
lock throughput on asymmetric processors~\cite{libasl}.

While algorithm customization can often resolve resource contention on
specific hardware, it may not utilize resources on other components
in a heterogeneous environment.
For example, adding backoff to locks on disaggregated 
memory can reduce communication resources usage on MNs, but the 
abundant communication resources on CNs remain underutilized.
Additionally, tuning lock algorithm parameters (e.g., the
backoff interval~\cite{backoff}) for varied hardware in
heterogeneous environments requires significant effort.

\stitle{Caching hot locks.}
Another approach is to cache frequently accessed locks on hardware
components with higher processing power or lower
communication costs to accelerate lock operations.
For example, deploying hot locks on programmable switches can
halve the network latency of distributed lock operations~\cite{netlock}.
Caching the lock in the local memory of each NUMA node
can improve overall lock throughput by minimizing expensive 
cross-NUMA memory accesses~\cite{cohort}.

However, simply caching locks on specific hardware components 
can lead to significant underutilization of their resources.
In the case of programmable switches, most of the switch memory is
allocated for storing holder and waiter information, 
which actually does not require fast processing.
Due to capacity limitations caused by memory shortage, even if
the cached locks are the most frequently accessed, only
a small fraction of lock requests are served by switches,
resulting in low utilization of their processing power~\cite{fisslock}.
Moreover, this approach cannot be generalized to heterogeneous 
hardware with weak processing power, e.g., SmartNICs.

\subsection{Motivating Case: Lock Fission}

Recent research has introduced a new lock design for
programmable networks known as lock fission~\cite{fisslock}. 
The core idea is to decouple the management of each lock between 
programmable switches and servers, 
leveraging the high processing power of switches alongside
the extensive memory capacity of servers. 
Specifically, the lock acquisition process is decomposed into
a latency-critical grant decision phase and 
a memory-intensive metadata maintenance phase.
The decision phase is handled by programmable switches
to minimize latency, while the maintenance phase is 
executed asynchronously by servers to conserve switch memory.
This approach enables programmable switches to accelerate operations 
on millions of locks, two orders of magnitude more 
than simply caching hot locks on the switch~\cite{netlock}.

%% file: proposal.tex
\section{The Principle}
\label{sec:principle}

Inspired by lock fission~\cite{fisslock}, we propose a new design principle 
for adapting locks to heterogeneous environments: \emph{the decomposition
of lock operations and metadata across various hardware components.}
By decomposing lock operations and metadata into independent parts that
align with the resource attributes of different hardware components,
we can leverage the strengths of each component 
while mitigating its bottlenecks, thereby optimizing overall performance.

However, applying this principle to a specific heterogeneous environment 
poses challenges due to a lack of generality, as it requires a thorough analysis 
of both the lock management process and the specific hardware attributes involved. 
Therefore, we seek a systematic approach that provides guidelines 
for decomposing lock operations and assigning them to appropriate hardware components, 
facilitating the design of efficient locking schemes on diverse platforms.

\section{The Proposal: Modularized Locks}
\label{sec:prop}

To systematically decompose locks, we go beyond simply splitting 
the lock manager into two parts~\cite{fisslock}. Instead, we fully
\emph{modularize} the lock manager into its smallest units and 
characterize the resource requirements for each module.
This approach eliminates the need to derive specific decomposition 
plans for different heterogeneous environments. Developers can simply 
assign lock modules to hardware components by matching each module's 
resource requirements with the hardware's attributes.
Modules on the same component can then be combined
to minimize inter-module communication.

\subsection{Lock Modules}
\label{sec:prop:module}

\begin{figure}[t]
  \begin{minipage}{1.\linewidth}
    \centering\includegraphics[scale=.52]{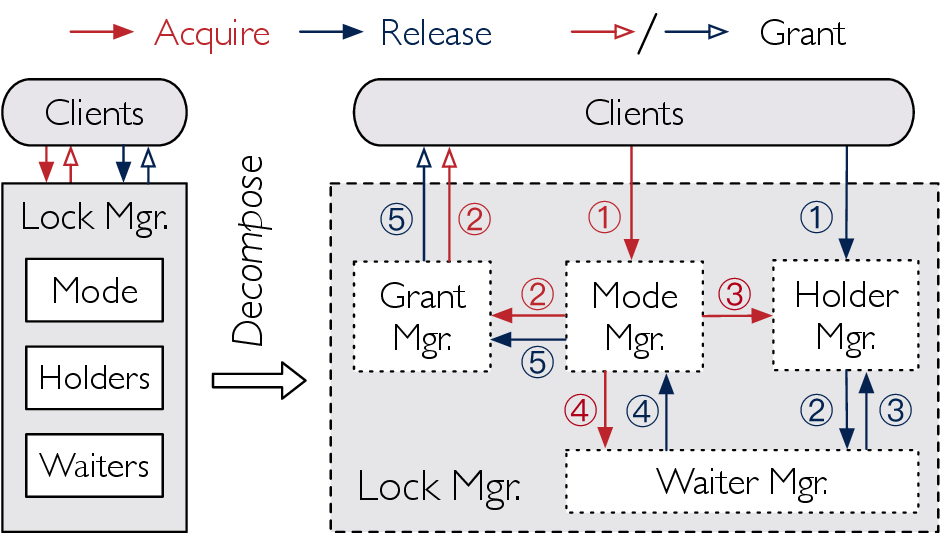}
  \end{minipage} \\[2pt]
  \begin{minipage}{1.\linewidth}
    \caption{\small{\emph{Architecture of traditional 
    lock managers (left) and modularized lock managers (right).}}}
    \label{fig:modules}
  \end{minipage}  \\[-5pt]
\end{figure}

As shown in Fig.~\ref{fig:modules}, we modularize the lock manager into 
four managers with distinct functionalities. Operations of each manager
are linearized using atomic operations or mutexes. 

\stitle{Mode manager.}
The mode manager maintains the essential lock states required
to determine whether a lock can be granted to a requesting client,
referred to as the \emph{lock mode}.
It uses minimal memory because the mode can typically be represented
with only a few bits.
However, it requires fast processing since it lies on the critical path 
of lock acquisition.

\stitle{Holder manager.}
The holder manager tracks the clients holding a lock (i.e., \emph{holders}).
For efficiency, it is often simplified to an atomic counter recording 
the number of holders~\cite{dslr,rdmarw}. 
However, in some cases, it is necessary to track metadata for each individual 
holder, such as determining whether their leases have expired~\cite{netlock}.
In these scenarios, the holder manager can require more processing and memory
resources.

\stitle{Waiter manager.}
The waiter manager tracks the clients waiting for a lock 
(i.e., \emph{waiters}) and determines which waiters should 
obtain lock ownership when the lock is released.
The resource requirements of the waiter manager
are determined by specific waiter selection policies.
Complex policies can guarantee powerful properties (e.g., fairness)
but require more processing and memory resources to 
enforce~\cite{netlock,fisslock,dslr}.

\stitle{Grant manager.}
The grant manager notifies clients when a lock is granted to them.
If the locations of waiting clients are known, the grant manager can 
send direct messages to them~\cite{netlock,fisslock}.
Otherwise, it relies on waiters to repeatedly check with the manager 
to see if the lock has been granted, which can be highly 
communication-intensive~\cite{dslr,rdmarw,netchain}.

\subsection{Lock Operations}
\label{sec:prop:op}

\nospacestitle{Lock acquisition.}
Clients acquire a lock by sending an acquisition request to the 
mode manager ({\ding[1.2]{192}}). 
The mode manager decides whether the lock can be granted based 
on the current lock mode and the mode of acquisition.
The result of acquisition is then communicated 
back to the clients through the grant manager ({\ding[1.2]{193}}).
If the lock is granted, the lock mode is updated to the mode
of acquisition, and the request is forwarded to the
holder manager, which updates holders to include the
requester ({\ding[1.2]{194}}). Otherwise, the request goes to the waiter
manager, which records the requester as a waiter ({\ding[1.2]{195}}).

\stitle{Lock release.}
Clients release a lock by sending a release request to the
holder manager, which removes the requester from holders 
({\ding[1.2]{192}}).
If no holder remains, the request is forwarded to the waiter
manager ({\ding[1.2]{193}}), which selects appropriate waiters to 
transfer lock ownership. After the selection, the waiter manager 
notifies the holder manager to record selected waiters as holders
({\ding[1.2]{194}}), and requests the mode manager to update the 
lock mode to the mode of selected waiters ({\ding[1.2]{195}}). 
After the update, the mode manager notifies selected waiters 
that the lock has been granted through the grant manager ({\ding[1.2]{196}}).

\subsection{Challenges}
\label{sec:prop:challenge}

\nospacestitle{Correctness.}
Although each modularized manager is linearizable, they are executed
in different orders during lock acquisition and release operations,
which may violate the linearizability of lock operations.
Thus, preserving the exclusion property
of locks after modularization is challenging. 

We propose a \emph{grant counting} method as a preliminary solution.
This method uses an optimistic design that forces the mode manager
to validate its grant decision before updating lock metadata and 
notifying clients. To achieve this,
the mode manager maintains a grant counter for each lock, recording 
how many times it has been granted to a client. The counter value
is synchronized to the holder manager when new holders
are added. Upon lock release, the holder manager embeds the grant 
count in the request forwarded to the waiter manager.
This grant count is then passed to the mode manager
when requesting a lock mode update. If the passed count
matches the count maintained by the mode manager, indicating 
no clients were granted the lock during the release, 
the mode manager updates the lock mode and informs the grant
manager to notify the selected waiters. Otherwise, 
the mode manager requests the holder manager and the waiter
manager to move selected waiters back from holders to waiters,
and aborts the grant process.

\stitle{Performance.}
Heterogeneous environments often include various hardware
components with differing resource attributes.
This diversity makes it challenging to devise a module assignment 
that maximizes the utilization of all components.
Although designing the optimal assignment for a specific environment
remains an open question, we offer some heuristics to guide the 
module assignment as follows.
We will showcase the effectiveness of these heuristics with
two examples in \S\ref{sec:example}.

\squishlist
\item \textbf{Prioritize managers on the lock granting critical path (H\#1)}.
The assignment should first meet the resource requirements of the mode 
and grant managers before others, ensuring optimal critical-path performance.

\item \textbf{Consider bottlenecks before strengths (H\#2)}.
Modules with high requirements for certain resources should not 
be assigned to components that are short of those resources,
as resource bottlenecks in heterogeneous hardware often 
overshadow performance gains brought by their strengths. 
\item \textbf{Fuse modules assigned to the same hardware (H\#3)}.
Lock modules are designed as the smallest units within lock 
managers to provide high flexibility. However, this introduces 
redundant interactions among modules on the same 
component, which can be avoided by fusing them.
\squishend

\begin{figure}[t]
  \vspace{1mm}
  \begin{minipage}{1.\linewidth}
    \centering\includegraphics[scale=.57]{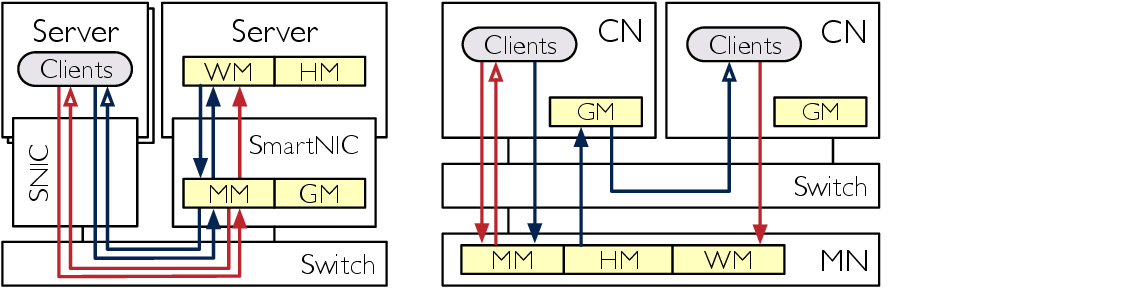}
  \end{minipage} \\[2pt]
  \begin{minipage}{1.\linewidth}
    \caption{\small{\emph{The application of modularized locks on
    SmartNICs (left) and disaggregated memory (right).
    }}}
    \label{fig:app}
  \end{minipage}  \\[-5pt]
\end{figure}

\section{Application Examples}
\label{sec:example}

We demonstrate the application of modularized locks in
two heterogeneous environments shown in Fig.~\ref{fig:app}.

\stitle{SmartNICs.}
In this example, each server is equipped with a SmartNIC, that
connects it to a top-of-rack switch. Application clients on multiple
servers acquire and release locks by sending requests to a lock service 
running on a dedicated lock server. In this scenario, a high rate of
lock requests can overwhelm the lock server's CPUs, causing significant
queueing delays. A strawman approach is to offload some locks to
datapath processors on the SmartNICs, alleviating the burden on CPUs. 
While these processors offer high parallelism
(e.g., hundreds of threads~\cite{bf3}) and shorten the network path,
they have limited cache sizes, typically 1\,KB per thread~\cite{dpu3study}. 
This leads to frequent cache misses when processing lock requests.
Microbenchmarks estimate that these cache misses can double the
request latency compared to using CPUs.

We use modularized locks to exploit the low network latency of
SmartNICs while minimizing the performance impact of cache misses.
Following H\#1, we assign the mode manager (MM) and the grant
manager (GM) to the SmartNIC to reduce network latency.
Both managers require minimal memory:
the MM accesses only a 2-bit lock mode, allowing thousands of 
locks to fit in each thread's L1 cache, and the GM uses only
a packet buffer for sending grant replies.
Following H\#2, the holder manager (HM) and the waiter manager 
(WM) are assigned to servers, preventing them from competing 
with the MM for the limited cache on SmartNICs.
Based on performance studies of SmartNICs~\cite{dpu3study},
we estimate that when lock conflicts are rare, this modularized 
design can reduce the network latency of lock acquisition by 
30\% and improve lock acquisition throughput by 5.8\x.

\stitle{Disaggregated memory.}
In this example, several CNs and MNs are connected via an RDMA network. 
Lock metadata is stored on MNs alongside application data.
Application clients running on CNs acquire locks and access data
by directly accessing MN memory via RDMA operations.
Due to the lack of computing power on MNs, the grant manager (GM) 
must rely on waiters repeatedly polling its state via RDMA 
to implement notifications.
As evaluated with a DM B$^+$-Tree index, with dozens of clients,
these frequent RDMA operations can saturate MNs' network resources,
leading to a throughput degradation of the index by more than 20\x.

Modularized locks help eliminate repeated RDMA 
operations to MNs during lock acquisitions. Following H\#1, we
allocate abundant communication resources to the GM by assigning it
to CNs. To facilitate this, we record the locations of waiters 
in the waiter manager (WM) on MNs. During lock release, after 
retrieving the waiters' location from the WM, the GM directly notifies 
waiters using the network between CNs. Following H\#3, managers on MNs
combine their metadata into a queue of holders and waiters,
enabling atomic acquisition and release with at most two RDMA
operations. In our experiments, this design reduces MN 
communication resource usage by over 95\% under high contention, 
boosting the throughput of the index by up to 27.7\x.

%% file: discussion.tex
\section{Next Steps}
\label{sec:future}

Our preliminary modularization of locks opens up numerous opportunities for future research. 
This section outlines several interesting directions to explore.

\stitle{Acquisitions from heterogeneous hardware.}
In the examples discussed in \S\ref{sec:example}, 
locks are acquired by clients operating on the same type of hardware.
However, this might not be the case in other heterogeneous environments.
For instance, in heterogeneous computing, CPUs and GPUs on the same SoC 
may need to synchronize the transmission of intermediate data 
by acquiring the same lock.
These scenarios require modularized locks that support 
different implementations for various hardware components, introducing new challenges in the assignment and implementation of lock modules.

\stitle{Dynamic assignment of modules.}
To accommodate workloads with dynamically changing patterns, 
a static assignment of modules may not sustain high utilization of heterogeneous resources.
This is primarily due to the varying resource requirements of lock modules 
under different workloads. 
For example, the memory consumption of the waiter manager is proportional to 
the number of waiters, which is influenced by the degree of lock contention.